# New structure of Channel Coding: Serial concatenation of Polar codes

Mohammed Mensouri and Mustapha Eddahibi

IMIS Laboratory, Ibn Zohr University, Agadir, Morocco

## Abstract

*In this paper, we introduce a new coding and decoding structure for enhancing the reliability and performance of polar codes, specifically at low error rates. We achieve this by concatenating two polar codes in series to create robust error-correcting codes. The primary objective here is to optimize the behavior of individual elementary codes within polar codes. In this structure, we incorporate interleaving, a technique that rearranges bits to maximize the separation between originally neighboring symbols. This rearrangement is instrumental in converting error clusters into distributed errors across the entire sequence. To evaluate their performance, we proposed to model a communication system with seven components: an information source, a channel encoder, a modulator, a channel, a demodulator, a channel decoder, and a destination. This work focuses on evaluating the bit error rate (BER) of codes for different block lengths and code rates. Next, we compare the bit error rate (BER) performance between our proposed method and polar codes.*

## Keywords

## 1. Introduction

Error correction codes are considered one of the most important elements of the 5G technology. The 5G standardization process selected two new channel coding schemes. Low Density Parity Codes (LDPC) have been adopted as a data coding scheme[1].They are designed to support high throughput, variables code rate and code length, in addition to very good error correction capability. Conversely, polar codes have been embraced as a novel coding technique for control information. Their design focuses on achieving effective error correction performance with concise block lengths and accommodating a wide range of code rates, all while maintaining stringent decoding latency requirements [2][3]. Polar codes are one of the latest additions to the family of error-correcting codes . They have been selected as the encoding method for the control channel in the fifth-generation cellular mobile communications network.

Large capacity, low latency, increased dependability, high data rates, and improved quality of services (QoS) are all features of the new 5G technology [4]. 5G technology is expected to significantly improve service quality given the volume of data on the network and the diversity of services. The performance seen by the user can be increased by reducing the BER (Bit Error Rate), i.e. by combining appropriate encoding and decoding schemes. When data is transmitted over a link, errors are introduced into the system. These errors reduce system performance. Therefore, the errors are calculated using BER. This value is calculated by dividing the number of bits received in error by the total number of bits transmitted during the same time period. This is the rate at which errors occur during transmission. Bit Error Rate (BER) and signal-to-noise ratio (SNR) are inversely proportional to each other. A BER of $10^{-9}$ is typically regarded as an acceptable level for telecommunications, whereas a minimum BER of $10^{-13}$ is considered more appropriate for data transmission.

In the present work, we harken back to the classical notion of code concatenation, a strategy that involves the sequential connection of two Polar Codes. These Polar Codes are interlinked in a series configuration, complemented by an interleaver structure. This interleaver orchestrates the permutation of bits within the outer codewords, and an inner encoder subsequently processes these permuted outer codewords, collectively contributing to the efficiency of the concatenated codes.

The structure of this paper is as follows. In Section 2, we will present the principles of concatenated codes, which form the foundation of our work. In Section 3, we describe the encoding and the decoding of Polar codes. The main contribution of the paper is Section 4, where we have developed a new coding and decoding structure which consists of concatenating two polar codes in series to build powerful error-correcting codes. Experimental results are given in section 5 that compares the error performance of serial concatenation Polar Codes for different block lengths and code rates. Finally, the conclusion is given in Section 6.

## 2. RELATED WORKS

The concept of concatenated codes, which encompass error-correcting codes constructed by amalgamating two or more simpler codes, represents a powerful approach that offers impressive performance while maintaining a manageable level of complexity. These codes were originally introduced by Forney in 1965 [5] as a solution to address theoretical challenges. By the 1970s, they had already found widespread utilization, particularly in the domain of space communications [6]. Subsequent developments in the realm of concatenated codes have given rise to the emergence of turbo codes and other modern, near-capacity codes, marking a significant evolution of this coding approach [7]. The fundamental premise behind code concatenation is to leverage the performance benefits of multiple codes, whether through a serial or parallel architecture, thereby ushering in a multitude of advanced code concatenation techniques that transcend the simplicity of a basic coding system.

One notable outcome of code concatenation is the potential to enhance the bit error probability. This enhancement stems from the synergistic combination of two or more codes, culminating in the creation of robust, higher block-length codes that are often accompanied by the incorporation of interleavers [8]. These interleavers serve to optimize the arrangement of bits within the codewords, facilitating reliable data transmission at rates that approach the channel capacity. It is important to note, however, that this gain in performance may be offset by an increase in decoder complexity.

Here are some common examples of code concatenation in the field of digital communications:

- Reed-Solomon and Convolutional Code Concatenation: As explained earlier, Reed-Solomon codes are often used as inner codes to correct random errors, while convolutional codes are used as outer codes to correct burst errors.
- LDPC and Convolutional Code Concatenation: LDPC (Low-Density Parity-Check) codes are error correction codes effective for burst error correction. They are often used in conjunction with convolutional codes to enhance error correction performance.
- Turbo Code: Turbo codes are a form of code concatenation. They consist of two separate convolutional codes with an interleaver in between. Data is first encoded by one convolutional code, interleaved, and then encoded again by another convolutional code. This significantly improves error correction performance.

These examples demonstrate how code concatenations are used to combine the advantages of different types of codes to achieve optimal error correction performance.

## 3. POLAR CODES

### 3.1. Polar Coding

Polar Codes are linear block codes. Their invention was proposed in [9]. Polar Codes apply channel polarization transformation to divide binary channels into perfect or completely noisy channels [2]. They then assign the information bits to the $K$ most reliable binary channels while the remaining bits are frozen, i.e., they are all set to a known value, usually "0". Similarly, for codeword length $N = 2^n, n \geq 1$, The polar code $(N, K)$ is characterized as a block code comprising $K$ input bits and $N$ output bits, featuring the generation matrix G, which is formed by taking the $n^{th}$ Kronecker power of the matrix F, as follows:

$$F = \begin{bmatrix} 1 & 0 \\ 1 & 1 \end{bmatrix} \qquad\qquad 1$$

$$G = F^{\otimes n} = \begin{bmatrix} F^{\otimes n-1} & 0_{n-1} \\ F^{\otimes n-1} & F^{\otimes n-1} \end{bmatrix} \qquad\qquad 2$$

The encoding procedure is executed through matrix multiplication, denoted as $X = U.\,G$, wherein $U$ represents the input vector sequence $U = (u_0, u_1, \ldots, u_{N-1})$ comprising both information bits and frozen bits, while $X$ signifies the resulting encoded vector $X = (x_0, x_1, \ldots, x_{N-1})$.

For example, the Polar Code (8,4) and $U = [0, 0, 0, u_3, 0, u_5, u_6, u_7]$ the corresponding codeword is:

$$X = U.\,G \qquad\qquad 3$$

$$X = [0, 0, 0, u_3, 0, u_5, u_6, u_7] \times \begin{bmatrix} 1 & 0 & 0 & 0 & 0 & 0 & 0 & 0 \\ 1 & 1 & 0 & 0 & 0 & 0 & 0 & 0 \\ 1 & 0 & 1 & 0 & 0 & 0 & 0 & 0 \\ 1 & 1 & 1 & 1 & 0 & 0 & 0 & 0 \\ 1 & 0 & 0 & 0 & 1 & 0 & 0 & 0 \\ 1 & 1 & 0 & 0 & 1 & 1 & 0 & 0 \\ 1 & 0 & 1 & 0 & 1 & 0 & 1 & 0 \\ 1 & 1 & 1 & 1 & 1 & 1 & 1 & 0 \end{bmatrix} \quad 4$$

$$X = \begin{bmatrix} u3 + u5 + u6 + u7 \\ u3 + u5 + u7 \\ u3 + u6 + u7 \\ u3 + u7 \\ u5 + u6 + u7 \\ u5 + u7 \\ u6 + u7 \\ u7 \end{bmatrix}$$

The codeword obtained by multiplying $U$ and $G$, as shown in this example, is non-systematic because information bits are not part of the codeword. A forward error correction code is considered systematic when it allows for a clear differentiation between information bits and parity bits. The key advantage of systematic codes lies in the straightforward appending of parity data to the source block. In cases where received data is correct, receivers are not required to retrieve the original source symbols. A block code can be represented in the form of a factor graph as explained in [10]. In the context of Polar Codes, we've noted that the generator matrix is constructed through a recursive process. It is then feasible to illustrate that the formation of the graph similarly follows a recursive pattern.

In this example, two generating matrices and their representation graph are shown. In figure 1 the generator matrix and the coding graph of a Polar Code of size $N = 2$ and message $U = [u_0, u_1]$ are presented.

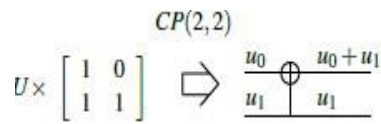

Figure 1. Generator matrix and graphical representation of the encoder for a Polar Code (2,2)

More generally, the factor graph of a Polar Code, as described in [11], with a size of $N = 2^n$, is structured into n stages. Each stage consists of N/2 parity nodes with a degree of 3 and N/2 variable nodes, also with a degree of 3. The degree of a node signifies the number of its connections to other nodes. The factor graph can be used for encoding and decoding. For coding, the input vector $U$, on the left side, is propagated in the graph in order to generate the code word X, on the right. An illustration of a Polar Code with an efficiency of $R = 0.5$ is provided in Figure 2. This implies that half of the vector $U$ comprises frozen bits, while the other half consists of information bits.

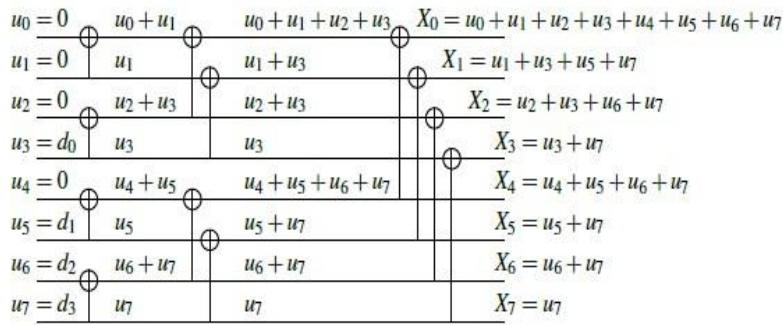

Figure 2 : Factor graph of a non-systematic CP(8;4) Polar Code encoder

## 3.2. Decoding Algorithms of Polar Codes

Decoding algorithms, as a general concept, represent a procedure through which the data received at the channel's output is analyzed and processed to retrieve the originally transmitted information while minimizing errors. The complexity, speed, and efficiency of decoding algorithms can vary significantly. Up to now, many efficient polar code decoding algorithms have been reported since the first polar code decoding algorithm was proposed in [2], Successive Cancellation decoder (SC). Although Successive Cancellation (SC) exhibits excellent performance with extremely long polar codes, it experiences a notable decline in maximum-likelihood decoding performance when dealing with short to medium block lengths. In an effort to address this problem, several variants of SC decoders, such as Successive Cancellation List (SCL) [11], Fli Successive Cancellation (SCF) [12], and Stack Successive Cancellation (SCS) [13], have been introduced, albeit at the expense of increased complexity.

Nonetheless, owing to the inherent serial processing of SC-based decoding algorithms, all of the previously mentioned approaches encounter notable drawbacks including substantial decoding latency and reduced throughput, thus impacting their practical usability. These algorithms yield hard outputs, meaning they produce binary results. On the other hand, fully parallel decoding algorithms, like the belief propagation (BP) method with soft output, have garnered considerable interest. The performance of BP decoding, based on Forney's factor graph representation, has been extensively examined in [14] [15].

The primary decoding algorithm for these codes is the hard decision method, commonly known as Successive Cancellation (SC). It stands as the most commonly utilized approach, and we will delve into its details in the following sections.

### 3.3. Decoding of the Polar Codes by Successive Cancellation

After the message has traversed the communication channel, the received noisy version $Y = (y_0, y_1, \ldots, y_{N-1}$ of code word $X = (x_0, x_1, \ldots, x_{N-1})$ is received. The aim of the decoding is to estimate the vector $U = (u_0, u_1, \ldots, u_{N-1})$ from the noisy version of the code word Y. In Arıkan's work presented in [9], it was demonstrated that Polar Codes can achieve the channel capacity when decoded using the successive cancellation algorithm. This decoding process involves the estimation of a bit $u_i$ based on observations from the channel and the information about previously estimated bits. The value of the estimated bit is denoted $\hat{u}_i$ with $U = (\hat{u}_0, \hat{u}_1, \ldots, \hat{u}_{N-1})$.

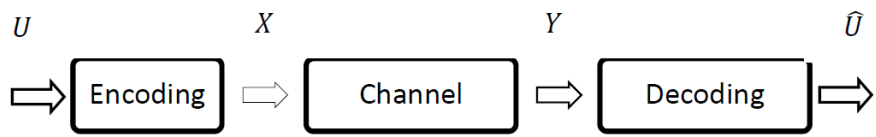

Figure 3. Error correcting communication over a noisy channel.

The decoding can be represented as a factorization graph. During the encoding process depicted in Figure 2, the information bits and the frozen bits are positioned on the left side of the graph. The progression from the left to the right of the graph is employed to compute the code word bits. During decoding, dual operation of encoding, the direction of the path is reversed. Channel information comes from the right (Figure 3). They represent the estimates of the bits of the code word. The steps for decoding the size 2 elementary kernel are detailed in Figure 2.4. The input data of the decoding algorithms presented here are therefore LLRs, contained in the vector $L$ of size $N$, taken out of the composite channel. The output data of the algorithm decoding are bits contained in the vector $U$ of size $K$. During the polar code decoding process, both data formats, LLRs and partial sums (PS), are used. The LLRs denoted $L_{i,j}$ are the estimates of the bit value at the position $(i, j)$ of the factorization graph. The partial sums noted $s_{i,j}$ correspond to the hard decision of this same bit of the graph of factorization.

The first step(a) of the decoding is the loading of the LLRs of the L-channel: LLR $L_{0,0}$ and $L_{1,0}$ take the value of the LLRs of the channel. The LLRs and partial sums of each point in the factorization graph are then calculated via the operations $f$, $R0$, g, $R1$ and $h$ symbolized in Figure 2.4 by arrows. They correspond to the following equations:

$$L_i = log\left(\frac{Pr\,(y_j|x_i=0)}{Pr\,(y_j|x_i=0)}\right) \qquad\qquad 5$$

$$f(L_a, L_b) = sign(L_a, L_b).\,min\,(|L_a|, |L_b|) \qquad 6$$

$$g(L_a, L_b, \hat{s}_a) = (1 - 2\,\hat{s}_a)L_a + L_b 7 h(\hat{s}_a, \hat{s}_b) = (\hat{s}_a \oplus \hat{s}_b, \hat{s}_b)\ 8$$

$$R0(L_a) = 0 \qquad\qquad\qquad\qquad 9$$

$$R1(L_a) = \begin{cases} 0 & if \quad L_a \geq 0 \\ 1 & if \quad L_a < 0 \end{cases} \qquad\qquad 10$$

The function f is applied in step(b). It allows the calculation of $L_{1,0}$. Step(c) is the application of the $R0$ function on the top node. The $R0$ function is applied assuming that the $u_0$ bit is a frozen bit. The function g then allows, in step (d), the calculation of $L_{1,1}$. Step (e) is the calculation of $s1;1$ by the corresponding R1 operation This operation is a thresholding of the LLR to obtain the partial sum. When the threshold is applied, the reliability information is lost. After decoding the partial sum $s_{1,1}$, the function $h$ is applied in order to propagate the partial sums, in step (f).

## 4. SERIAL CONTINUATION OF POLARCODES

### 4.1. Encoding

A serial concatenation of Polar Codes is obtained by the combination of two Polar codes in a serial way and they are separated by interlevaer $\pi$ as shown in figure 4. The first polar code C1 is called external code, while the second polare code C2 is internal code. The information bits are encoded by the external code, interleaved through an interleaver ($\pi$) and then re-encoded by the internal code. The interleaver rearranges the order of the bits transmitted, while the corresponding deinterleaver restores the original order. Deinterlacing can thus disperse a burst of errors associated with isolated errors so that these individual errors may be easier to correct by the second decoder.

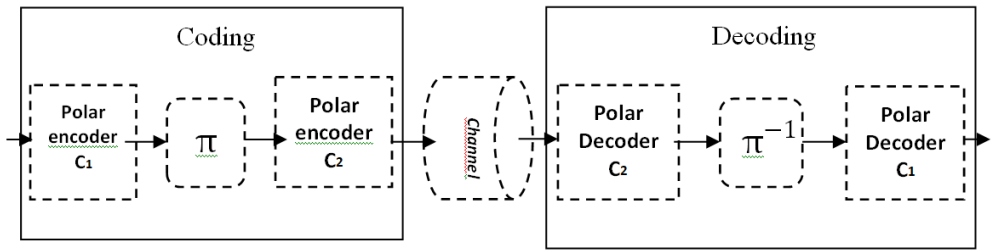

Figure 4. Series concatenation of polar codes

Considering that $R_1 = {}_N{}^K 1_1$ and $R_1 = {}_N{}^K 2_2$ are the respective rate of the polar code C1 and the polar code C2. The overall rate of the code built is directly related to the rate $R1$ and $R2$ of the two Polar Codes C1 and Polar Codes C2 by the relation:

$$R_s = R_1 \times R_2 \qquad\qquad 11$$

If $R1$ and $R2$ are identical, the preceding relation becomes:

$$R_s = R_1{}^2 \qquad\qquad 12$$

## 4.2. Decoding the Concatenated Polar Codes

At the decoding level, there are two Polar Decoders 1 and Polar Decoders 2 concatenated in series and separated by a deinterleaver $\pi^{-1}$ as shown in figure 4 part decoding. In fact, decoder 2 receives as input the sequences generated by encoder 2. The Polar decoder 2 generates at output a sequence which has many grouped errors. The purpose of the deinterleaver $\pi^{-1}$ then is to disperse these errors to increase the correction efficiency of the Polar decoder 1.

## 4.3. Interleaver and Deinterleaver

Interleaving involves reorganizing a sequence of bits in a way that maximizes the separation between symbols that were initially located close to one another, as explained in reference [16]. This makes it possible in particular to transform an error relating to grouped bits into an error distributed over the whole of the sequence. He There are several types of interleavers. An interleaver is a device that reorders a data sequence using a deterministic bijective mapping. Let $C = (c_0, c_1, \ldots, c_{N-1})$ be a sequence of length N. An interleaver transforms $C$ into a sequence $X = (x_0, x_1, \ldots, x_{N-1})$ in such a way that $X$ represents a permutation of the elements in $C$. When we treat $C$ and $X$ as a pair of N-dimensional vectors, there is a one-to-one correspondence, denoted as $c_i \rightarrow x_j$, between each element in $C$ and each element in $X$, as illustrated in Figure 5.

Let $I = \{0, 1 \ldots N-1\}$. An interleaving operation can be characterized through the one-to-one index mapping function:

$$\pi: \quad I \rightarrow I \qquad\qquad 13$$
$$i \rightarrow j = \pi(i)$$

In this context, "$i$" denotes the index of an element within the original sequence $C$, while "$j$" corresponds to the index of the corresponding element in the interleaved sequence $X$. This mapping function can be represented as an ordered set known as the interleaving vector $\pi = (\pi(0), \pi(1), \ldots, \pi(N-1))$.. The $k^{th}$ element of the permuted sequence $X$ is given by:

$$x_k = c_{\pi(k)} \qquad\qquad 14$$

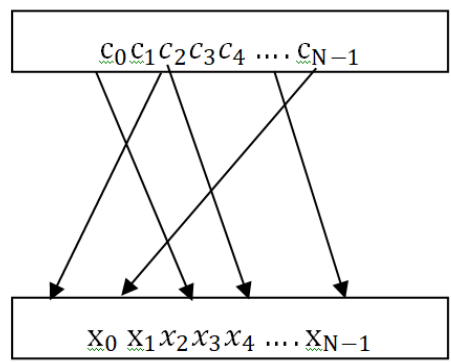

Figure 5. Mechanism of data interleaving

The deinterleaver, often referred to as the inverse interleaver, reverses the permutation, restoring the permuted sequence to its original order. In this paper, we employ $\pi$ to represent the interleaving vector and $\pi^{-1}$ for the deinterleaving vector. With the appropriate deinterleaver, the permuted elements can be repositioned to their original locations:

$$\pi^{-1}: \quad I \to I$$
$$j \to i = \pi^{-1}(j) \qquad\qquad 15$$

$$\pi^{-1}(\pi(k)) = \pi(\pi^{-1}(k)) = k \text{ If} \qquad\qquad 16$$

we replace $k$ by $\pi^{-1}(n)$ in equation 14, we get:

$$x_{\pi^{-1}(n)} = c_{\pi(\pi^{-1}(n))} = c_n \qquad\qquad 17$$

## 5. PERFORMANCE EVALUATION

In the current section, several performance curves are presented. These curves show the impact of parameters on decoding performance of the concatenated coding scheme described in the previous subsection for serial continuation of Polar codes . These parameters and their effects are important, as they are the levers used in communication standards to adapt to system constraints. The axes of the various curves are always the same. On the x-axis is the signal-to-noise ratio, Eb/N0. On the y-axis is the BER (Bit Error Rate). The simulations are executed to evaluate the system's behavior when data is transmitted through an AWGN channel, making use of Binary Phase Shift Keying (BPSK) modulation.

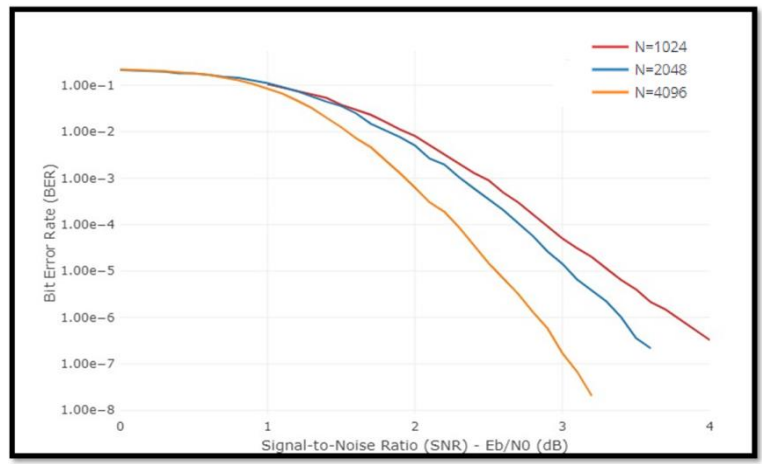

Figure 6. Decoding performance of the concatenated Polar Codes for different values of N With
$$R = 1/2$$

Figure 6 illustrates the decoding performance of concatenated Polar codes as codeword size increases. While it is clear that the bit error rate decreases for a given signal-to-noise ratio, the Shannon limit is a little far away for the codeword sizes shown.

Error-correcting codes are based on the notion of redundancy, quantified by the performance of the code. The lower the efficiency, the higher the redundancy. This is illustrated in Figure 7, again considering the SC algorithm. The observed trend is that the lower the yield, the lower the error rate. Figure 7 shows the performance of theconcatenated Polar Codes. The code considered is a code with a codeword size of 2048.

Figure 8 allows to compare the performances in BER between our proposed method and the polar code. According to this figure, we can see that the BER of our method is better than the simple polar code. Note that the length of the information sequence is K = 1723 bits and N = 2048 to do this simulation.

In this section, we discussed the simulation results of our method for different parameters. In the future works, we will compare the simulation results of our method against various errorcorrecting codes, such as LDPC and Turbo Codes.

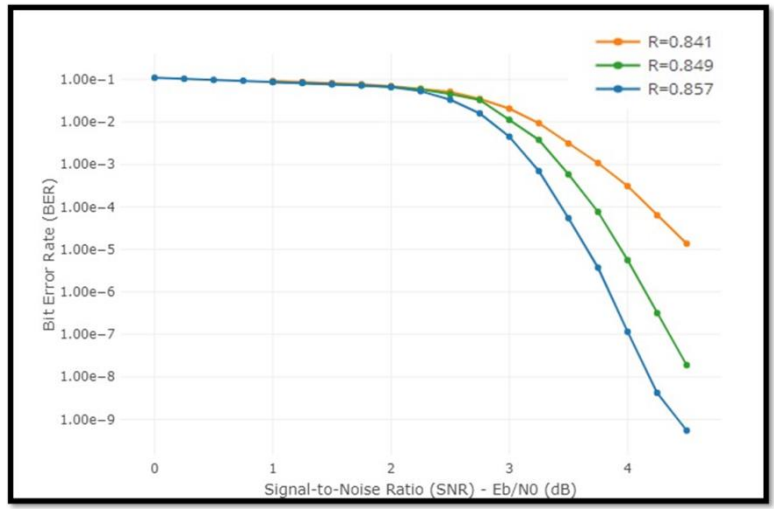

Figure 7. Decoding performance of theconcatenated Polar Codes for different values of R With  N= 2048

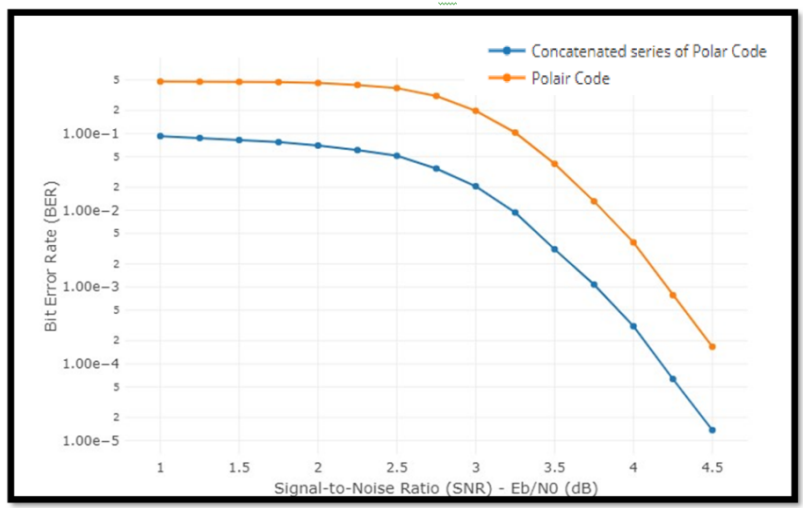

Figure 8. Comparison of decoding performance between concatenated series of Polar codes and Polar code with N= 2048

# 6. CONCLUSION

In this paper, we provide a comprehensive overview of polar codes, including their construction. Then, we presented the main algorithms for decoding polar codes. The decoding algorithm used, called SC, was detailed because it was used in the new encoding and decoding scheme. Through the various simulations presented in this paper, we noticed that this new coding and decoding structure by using polar codes has good performance of point of view BER. This structure consists of two concatenated polar codes in series, separated by an interleaver, in order to develop a highly efficient code suitable for use in 5G technology. The interleaver and deinterleaver play crucial roles in this coding and decoding structure. A well-designed permutation should enable us to achieve good performance in terms of Bit Error Rate (BER). Investigating the impact of the interleaver and deinterleaver on this structure will be the focus of future work.

## ACKNOWLEDGEMENTS

We would like to thank all authors for their contributions and the success of this manuscript and all editors and anonymous reviewers of this manuscript.

## REFERENCES

[1]    R.G. Gallager (1963), Low Density Parity Check Code, MIT Press, Combridge.
[2]    E. Arikan, (2009) "Channel polarization: A method for constructing capacity achieving codes for symmetric binary-input memoryless channels", *IEEE Trans. Information Theory*, Vol. 55, No. 7, pp. 3051–3073.
[3]    Y. Fan et al., (2015) "Low-latency list decoding of polar codes with double thresholding," *2015 IEEE International Conference on Acoustics, Speech and Signal Processing (ICASSP*), South Brisbane, QLD, Australia, pp. 1042-1046, doi: 10.1109/ICASSP.2015.7178128
[4]    Hani Attar, Haitham Issa, Jafar Ababneh, Mahdi Abbasi, Ahmed A. A. Solyman, Mohammad Khosravi, Ramy Said Agieb, (2022) "5G System Overview for Ongoing Smart Applications: Structure, Requirements, and Specifications", Computational Intelligence and Neuroscience, vol. 2022, pages 11. https://doi.org/10.1155/2022/2476841
[5]    G. D. Forney, Jr, (1966) *Concatenated Codes*, MIT Press, Cambridge, MA.
[6]    D. J. Costello & G. D. Forney, (2007 ) "Channel coding: The road to channel capacity," in *Proceedings of the IEEE*, Vol. 95, No. 6, pp. 1150-1177. Doi: 10.1109/JPROC.2007.895188.
[7]    C.Berrou, A.Glavieux, P.Thitimajshima, (1993)    "Near shannon limit errorcorrecting coding anddecoding: turbo-codes", *in: IEEE International Conferenceon Communication*, pp.1064– 1070.
[8]    H, P. ao, D., & Hoeher (2008), "Helical interleaver set design for interleave-division multiplexing and related techniques", IEEE Communications Letters, 12(11), pp. 843–845. doi:10.1109/LCOMM.2008.080990
[9]    E.Arikan (2008) "Channel polarization: a method for constructing capacity-achieving codes", *In IEEE International Symposium on Information Theory* (ISIT 2008), pp. 1173–1177.
[10]   H. Aurora, C. Condo & W. J. Gross (2018) "Low-Complexity Software Stack Decoding of Polar Codes", In IEEE International Symposium on Circuits and Systems (ISCAS), pp. 1–5.
[11]   I. Tal & A. Vardy , (2011 ) "List Decoding of Polar Codes", *In IEEE International Symposium on Information Theory (ISIT)*, pages 1–5.
[12]   Orion Afisiadis, Alexios Balatsoukas-Stimming & Andreas Burg, (2014) "A low-complexity improved successive cancellation decoder for polar codes", *48th Asilomar Conference on Signals, Systems and Computers, IEEE*, pp. 2116–2120.
[13]   Kai Niu and Kai Chen, (2012) "Stack decoding of polar codes", Electronics letters 48.12, pp. 695– 697.
[14]   Erdal Arıkan ,(2010) "Polar codes: A pipelined implementation", Proc. 4th ISBC, pp. 11–14.

[15]  Nadine Hussami, Satish Babu Korada, & Rudiger Urbanke (2009) "Performance of polar codes for channel and source coding" IEEE International Symposium on Information Theory, IEEE, pp. 1488–1492.

[16]  C. Berrou, Y. Saouter, C. Douillard, S. Kerouedan,& M. Jezequel , (2004) "Designing good permutations for turbo codes : towards a single model", *in IEEE International Conference on Communications (ICC),* vol. 1, pp. 341–345.

# AUTHORS

**Prof. Mohammed Mensouri** received the M.S degree in Networks and Telecommunication in 2008, from Faculty of Sciences and Technology, Cadi Ayyad University, Marrakech, Morocco. In 2015, He received Ph.D of Computer Science in Faculty of Sciences, Chouaib Doukkali University, El Jadida, Morocco.  He is professor at Ibn Zohr University, Agadir, Moroc. His research interest information theory and channel coding , especially error correction codes.

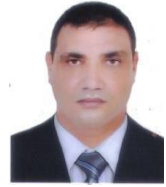

**Prof. Mustapha Eddahibi** received his PhD from Unversity Cadi Ayyad Marrakech in 2007. He is currently computer science teacher researcher in University Ibn Zohr. He is a former head of the decisional expert systems research team. His research interests are in the area of intelligent computing, information engineering, digital Information Encoding and Processing.

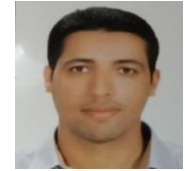